\documentclass[aps,preprintnumbers,amsmath,amssymb,floatfix,showpacs,
twocolumn]{revtex4}

\usepackage{epsfig} \usepackage{bm}

\newcommand{\kpm}{K}
\renewcommand{\H}{H$_2$~}
  
  \newcommand{\Hp}{H$_2^+$~}
\newcommand{\ba}{\begin{eqnarray}}
\newcommand{\ea}{\end{eqnarray}}
\newcommand{\br}{\begin{eqnarray*}}
\newcommand{\er}{\end{eqnarray*}}
\newcommand{\be}{\begin{equation}}
\newcommand{\ee}{\end{equation}}
\newcommand{\eref}[1] {(\ref{#1})}
\newcommand{\Eref}[1] {Eq.~(\ref{#1})}
\newcommand{\Fref}[1] {Fig. \ref{#1}}

\begin{document}

\title{The $p$--H symmetry breaking in dissociative ionization of H$_2$ due to
the molecular ion interaction with the ejected electron}

\author{Vladislav V. Serov}
\affiliation{
Department of Theoretical Physics, Saratov State University, 83
Astrakhanskaya, Saratov 410012, Russia}

\author{ A. S. Kheifets}

\affiliation{Research School of Physical Sciences,
The Australian National University,
Canberra ACT 0200, Australia}

\date{\today}

\begin{abstract}
We propose a mechanism of electron localization and molecular symmetry
breaking in dissociative photoionization of the \H molecule. The
Coulomb field of the ejected electron can induce transition of the
remaining H$_2^+$ ion from the gerade $^2\Sigma_g^1(1s\sigma_g)$ to
the ungerade $^2\Sigma_u^1(2p\sigma_u)$ electronic state when the
nuclei in a bound vibrational state are near the outer turning
point. The superposition of this process with a direct transition to
vibrational continuum should produce a non-gerade ionic state which
results in observed asymmetry in the $p$--H ejection relative to the
electron ejection direction at a small kinetic energy release.
\end{abstract}



\maketitle


Dissociative photoionization (DPI) of the hydrogen molecule has been a
subject of considerable interest in recent years. Even though it is a
relatively weak single photoionization channel, the breakup of the \Hp
ion into the ionic H$^+\equiv p$ and neutral H atomic fragments allows
for determination of the molecular axis orientation when the reaction
products are detected in coincidence. Because the dissociation is fast
compared with molecular rotation, the direction of fragmentation
coincides with the molecular axis at the instant of
photoionization. This allows for the molecular frame photoelectron
angular distribution to be determined.

One of striking observations following from these coincident studies
was breakup of the photoelectron emission symmetry with respect to the
ionic $p$ and neutral H atomic fragments. This asymmetry was found in
single-photon induced DPI process
\cite{0953-4075-36-23-007,Martin02022007,0953-4075-45-19-194013,
PhysRevLett.110.213002} as well as in multi-photon regime
\cite{Kling14042006,PhysRevLett.103.213003,Sansone2010,PhysRevLett.104.023001,
PhysRevLett.107.043002,Wu2013}.  The finding of this asymmetry was
indeed surprising. The two single photoionization channels lead from
the ground \H state to the ionic gerade $^2\Sigma_g^1(1s\sigma_g)$ and
ungerade $^2\Sigma_u^1(2p\sigma_u)$ states that are well separated in
energy and do not normally mix. Hence, the ionic state possesses a
well-defined exchange symmetry which leads to the fully symmetric
photoelectron emission.  The $p$--\,H asymmetry in dissociation means
that the asymptotics of the wave function of the state with a fixed
ejected electron energy and a fixed nuclear kinetic energy release
(KER) is non-gerade by the coordinates of the bound electron. From the
mathematical point of view, this means that the bound electron is in a
superposition of gerade and ungerade \Hp wave functions.

It was Mart\'{\i}n and Fern\'andez
\cite{Martin02022007,1367-2630-11-4-043020} who showed that the two
ionic states can actually mix because of autoionization via the $Q_1$
and $Q_2$ doubly excited states.  This creates a mixed parity ionic
state that allows for the localization of the bound electron and
angular asymmetry of the photoelectron. Since the autoionizing states
lives few femtoseconds, the nuclei have enough time to move outside
the Franck-Condon (FC) region before the electron is
ejected. Theoretical description of such a process therefore requires
going beyond the fixed nuclei Born-Oppenheimer approximation.

An additional mechanism that can be responsible for the photoemission
symmetry breaking is the effect of an intense IR field from few-cycle
laser pulses \cite{Kling14042006,Sansone2010,Wu2013}. The strong laser
field may act in conjunction with quantum mechanical interference
involving autoionizing states and the laser-altered wave function of
the departing electron. Alternatively, the charge localization may be
due to laser-driven transition between different electronic states of
the molecular ion.

An alternative mechanism that could be responsible for the DPI
asymmetry was discussed briefly by \citet{Martin02022007} in the form
of a preferred attractive interaction between the proton and the
escaping electron. However, the photoelectron is too fast to be
efficiently perturbed by the slow proton, except possibly in the
region of the maximum allowed KER. In the present work, we consider a
similar mechanism of direct photoelectron interaction with the
remaining ion.  The Coulomb field of the ejected electron can induce
transition of the remaining H$_2^+$ ion from the gerade
$^2\Sigma_g^1(1s\sigma_g)$ to the ungerade $^2\Sigma_u^1(2p\sigma_u)$
electronic state when the nuclei in a bound vibrational state are near
the outer turning point. The superposition of this process with a
direct transition to vibrational continuum should produce asymmetry in
the $p$--H ejection relative to the electron ejection direction at a
small kinetic energy release.


The $p$--\,H asymmetry in DPI means that the wave function of a state
with a fixed ejected electron energy and a fixed nuclear KER is
non-gerade by coordinates of the bound electron.  From the
mathematical point of view, this means that the bound electron is in
superposition of gerade and un-gerade H$_2^+$ wave functions.
The amplitudes of generation of the gerade and ungerade ionic states are
proportional to the corresponding FC factors states
\be
S_{g\kpm}=\langle g\kpm |\text{H$_2$}\rangle
\ , \ \ \ 
S_{u\kpm}=\langle u\kpm |\text{H$_2$}\rangle
 \ .
\ee
Here $|\text{H$_2$}\rangle \equiv \chi_{\text{H$_2$}}(R)$ is the
ground vibrational state of H$_2$, $|g\kpm\rangle \equiv \chi_{\kpm
L}^{1s\sigma_g}(R)$ and $|u\kpm\rangle \equiv \chi_{\kpm
L}^{2p\sigma_u}(R)$ are vibrational continuum functions of H$_2^+$ in
gerade and ungerade electronic states, respectively, $\kpm=\sqrt{2m
E_R}$ is the relative momentum of ion and atom after dissociation,
$E_R$ is the kinetic energy release, $m$ is the reduced mass of a
nucleus, $L$ is the angular momentum transferred to the rest ion in a
ionization.

\begin{figure}[ht] 
\includegraphics[angle=-90,width=0.9\columnwidth]{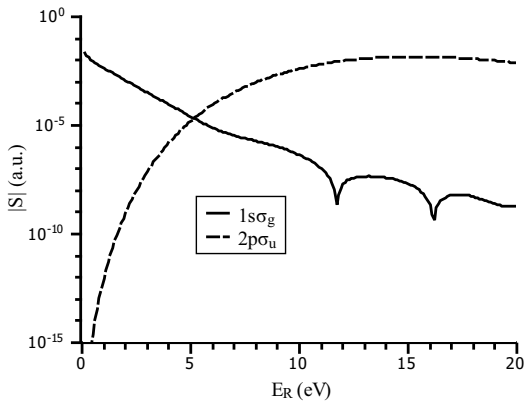}
\caption{Absolute values of the overlap integrals between the H$_2$
vibration ground state wave function and the H$_2^+$ vibration
continuum for the $1s\sigma_g$ (solid line) and $2p\sigma_u$ (dashed
line) electronic states.}
\label{FIG_Tp}
\end{figure}

The DPI amplitude via the $2p\sigma_u$ state for $E_R=1$~eV is lesser
then that via the $1s\sigma_g$ state by 10 orders of magnitude (see
\Fref{FIG_Tp}). Therefore, for small KER, a direct transition to an
ungerade state cannot induce an observable asymmetry. It is also clear
from a classical physics consideration. The turning point in the
$1s\sigma_g$ adiabatic potential for $E_R=1$~eV is $R=1.023$~a.u. The
turning point in the $2p\sigma_u$ adiabatic potential at the same KER
is $R=4.45$~a.u., well outside the FC region and where the H$_2$
ground vibration state wave function is extremely small.

The overlap integrals are equal for the gerade and ungerade ionic
states at $E_R=5$~eV. So, the asymmetry of the proton-atom ejection
for $E_R=5$~eV should be similar to asymmetry of the bound electron
localization immediately after the photoelectron ejection. But the
probability of DPI for $E_R=5$~eV is 4 orders of magnitude less then
for $E_R=1$~eV. Nearly the same value of KER appears in the case of
DPI via a quasistationary state, which can also lead to a strong
asymmetry \cite{Martin02022007,1367-2630-11-4-043020}.


As it is clear from the above arguments, DPI with a small KER proceeds
solely via the H$_2^+$ ground state. In result, the original asymmetry
in the bound electron wave function is lost entirely.
To make a transition to a vibrational continuum of $2p\sigma_u$ with a
small KER, the ion should reach an internuclear distance close to the
turning point in the $2p\sigma_u$ adiabatic potential for a small
above-threshold energy. After that, the H$_2^+$ ion should interact
with some external field, which can excite it to the $2p\sigma_u$
state. There are a number of experiments in which an IR laser field
with a small energy is used for this purpose
\cite{Kling14042006,Sansone2010,Wu2013}.

\begin{figure}[ht] 
\includegraphics[angle=-90,width=0.9\columnwidth]{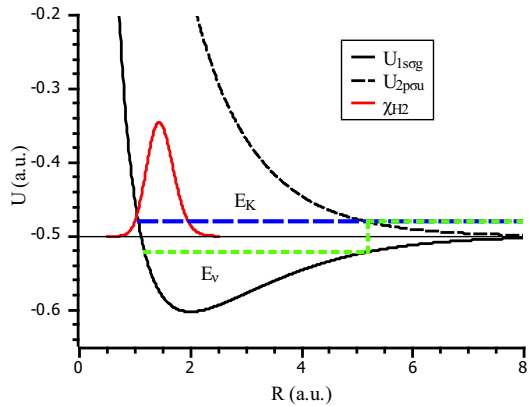}
\caption{(Color online) The energy diagram of the proposed DPI process. Exhibited in the figure are the adiabatic potentials for the ionic gerade $1s\sigma_g$ (solid line) and ungerade $2p\sigma_u$ (dashed line) states, the initial \H vibration state wave function (red solid line), the vibrational continuum energy level and vibrational bound energy level, and the two pathways of dissociation --- the direct transition to the continuum energy level $E_K$ (dashed blue line) and via an intermediate vibrational bound state $E_\nu$ with following transition of the ion to the $2p\sigma_u$ state (dotted green line).}
\label{FIG_ways}
\end{figure}

One should note, however, that in the DPI process of \H there is
always present an external field that interacts with the \Hp ion.  It
is the field of the ejected photoelectron. During the time while the
inter-nuclear separation reaches sufficient distance to make a
transition to a small-KER ungerade state, the ejected electron have
already flown far away, its interaction with the ion is
small. However, several effects can conspire to help the asymmetry
from interaction with far ejected electron to be observable:

\begin{enumerate}

 \item When the ion is in the bound vibrational state with the outer
 turning point close to inner turning point in the $2p\sigma_u$
 adiabatic potential (see \Fref{FIG_ways}), the nuclei spend a long
 time near the point $R$ were the overlap between vibrational states
 is maximal.

 \item The number of ions in bound vibrational states after ionization
 is much larger then ones in vibrational continuum states, and even a
 small conversion to ungerade dissociative state can give result
 comparable to a direct transfer to a gerade vibrational continuum
 state.
 
 \item The $1s\sigma_g$ and $2p\sigma_u$ states form pair of so-called
 charge-resonance (CR) states. The energy gap between the CR states is
 decreasing with the internuclear distance $R$, while the dipole
 transition matrix element $\langle 2p\sigma_u |z| 1s\sigma_g\rangle \to R/2$.  Thereby, CR states are strongly coupled to external field for large $R$.

\end{enumerate}

Let us estimate the asymmetry due to this process.
We assume that a duration of the ionizing pulse is much smaller then
the characteristic time of the nuclear motion. After ejection of the
photoelectron, the evolution of the vibrational state can be described
by a time-dependent Schr\"odinger equation
\be 
i\frac{\partial\chi_g(R,t)}{\partial
t}=\left[-\frac{1}{2m}\frac{\partial^2}{\partial
R^2}+\frac{L(L+1)}{2mR^2}+U_{1s\sigma_g}(R)\right]\chi_g(R,t)
\ ,
\label{gTDSE} 
\ee
where $U_{1s\sigma_g}(R)$ is the adiabatic potential. The initial
condition imposed on \Eref{gTDSE} is that its solution coincides with
the H$_2$ vibrational ground state
\be
\chi_g(R,0)=\chi_{\text{H$_2$}}(R)
\ .
\ee
The matrix element of transition between the gerade and ungerade ionic
states in the field of the ejected electron is given by the following
expression
\be
\mu(R,r_e)\simeq\frac{d(R)}{r_e^2}
\ .
\label{pert}
\ee
Here $r_e$ is the distance to the ejected electron and
\be
d(R)=\langle 2p\sigma_u |\mathbf{n}_e\cdot\mathbf{r}| 1s\sigma_g\rangle 
\label{dipoltr}
\ee
is the dipole matrix element, $\mathbf{r}$ is a coordinate of the bound
electron, $\mathbf{n}_e=\mathbf{r}_e/r_e$,
$|1s\sigma_g\rangle\equiv\varphi_{1s\sigma_g}(\mathbf{r};R)$,
$|2p\sigma_u\rangle\equiv\varphi_{2p\sigma_u}(\mathbf{r};R)$.
\Eref{pert} is derived by the Taylor's expansion of
\(V(\mathbf{r},\mathbf{r}_e)=1/|\mathbf{r}-\mathbf{r}_e|\). In a
case $R\gg 1$, the dipole matrix element $d(R)\approx \mathbf{n}_e\cdot\mathbf{R}/2$.

As we expect that the ejected electron is far from the ion at the time
of transition, we can describe this electron quasi-classically and
assign to it a trajectory $r_e(t)$.  Under this assumption, the
amplitude of transition to the ungerade state with the asymptotic
relative nuclear momentum $\kpm$ can be written as
\be
A_{u\kpm}=-i\int_0^\infty\frac{1}{r_e^2(t)}\langle u\kpm
|d(R)|\chi_g(R,t)\rangle e^{iE_\kpm t}dt
\ .  
\label{AUK}
\ee 
Here $|u\kpm\rangle\equiv\chi_{\kpm L}^{2p\sigma_u}(R)$ is a solution
of the stationary Schr\"odinger equation
\ba \left[-\frac{1}{2m}\frac{\partial^2}{\partial
R^2}+\frac{L(L+1)}{2mR^2}+U_{2p\sigma_u}(R)\right]\chi_{\kpm
L}^{2p\sigma_u}(R)&=& \nonumber \\ E_\kpm\chi_{\kpm L}^{2p\sigma_u}(R)
\ ,
\label{uSSE} 
\ea 
with $E_\kpm=E_R+U_{2p\sigma_u}(\infty)$.
The asymptotic form of the continuum function is $\chi_{\kpm
L}^{2p\sigma_u}(R\to\infty)=\sin(\kpm R+\delta_u)$, where $\delta_u$
is a scattering phase.
For all the results shown below, the wave functions were obtained by
numerical solution of Eqs.~\eref{gTDSE} and \eref{uSSE}.

The results are weakly depended on the angular momentum $L$, while $L$
is not very large. Typically $L$ is small when the $H_2$ molecule is
initially in a low rotational state and the angular momentum exchange
due to recoil of the ejected electron is also small. The latter
condition is always satisfied when the ejected electron has low
energy. For this reasons, the angular momentum $L=1$ (that corresponds
to ground rotational state of the ortho-hydrogen) was assumed in all
examples shown below.
 
With the simplest approximation $r_e(t)=v_e t$, where
$v_e=\sqrt{2E_e}$, $E_e$ being the ejected electron energy, the
equation for the amplitude \eref{AUK} can be rewritten as
\be A_{u\kpm}=-\frac{i}{E_e}\int_0^\infty a(t) dt \label{A_int_a} 
\ ,
\ee
where
\be 
a(t)=\frac{1}{2t^2}\langle u\kpm |d(R)|\chi_g(R,t)\rangle e^{iE_\kpm t}
\label{att} 
\ee 
is the amplitude of transition per unit time.

\begin{figure}[ht] 
\includegraphics[angle=-90,width=0.9\columnwidth]{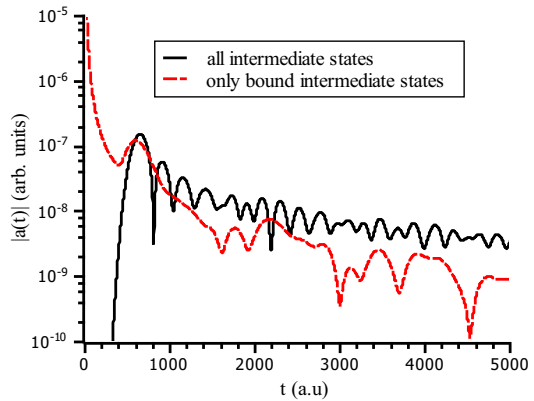}
\caption{(Color online) The absolute value $|a(t)|$ of the amplitude of transition
per unit time for $E_R=0.4$ eV from \Eref{att} (solid line) and
\Eref{attt} (dashed line).}
\label{FIG_at}
\end{figure}

The modulus of this amplitude is shown in \Fref{FIG_at} for selected
value of $E_R=0.4$~eV. \Fref{FIG_at} supports the qualitative
reasoning given above. The noticeable transition begins only when the
vibrational wave packet have spread up to the turning point
$R(E_R)=5.49$ a.u. corresponding to $U_{2p\sigma_u}(R)=E_K$. The peak
of transition is near $t\approx 700$ a.u. The ejected electron with an
energy, say, $E_e=2$ eV, at this time would reach the distance
$r_e\approx 250$ a.u.

It may seem that since the intermediate bound vibrational states
contribute predominately to the process under consideration, we can
take into account these states only in our consideration. In this
approximation, the solution of \Eref{gTDSE} can be written as
\be
\tilde{\chi}_g(R,t)=\sum_\nu \langle g\nu |\text{H$_2$}\rangle \chi_{\nu
L}^{1s\sigma_g}(R) e^{-iE_{\nu L}t} \ .
\label{SUM}
\ee
The amplitude \eqref{att} of transition per unit time takes the form
\be
\tilde{a}(t)=\frac{1}{2t^2}\sum_\nu\langle u\kpm |d(R)|g\nu\rangle \langle g\nu
|\text{H$_2$}\rangle e^{i\omega_{\kpm\nu}t} \ ,
\label{attt}
\ee
where $\omega_{\kpm\nu}=E_\kpm-E_{\nu L}$.  
However, such approach is unapplicable. In \Fref{FIG_at}, we compare
the moduli of the amplitudes $a(t)$ from \Eref{att} and $\tilde{a}(t)$
from \Eref{attt}. From this comparison, we see that omitting of
continuum from the summation over the intermediate vibronic states
causes a serious error. While the position and shape of the main peak
does not differ significantly, the amplitude $\tilde{a}(t)$ has an
unphysical peak at small $t$ and diverges at $t\to 0$. This peak
appears because the omitting of the continuum delocalizes the initial
wave packet. That is why in a calculation of $A_{u\kpm}$ we used the
direct solution of \Eref{gTDSE} instead of approximated analytical
solution given by \Eref{SUM}.

The asymmetry in the $p$--H ejection can be expressed as \cite{Wu2013}
\ba
\beta &=&
\frac{|S_{g\kpm}e^{i\delta_g}-A_{u\kpm}e^{i\delta_u}|^2-|S_{g\kpm}e^{i\delta_g}+A_{u\kpm}e^{i\delta_u}|^2}{|S_{g\kpm}e^{i\delta_g}+A_{u\kpm}e^{i\delta_u}|^2+|S_{g\kpm}e^{i\delta_g}-A_{u\kpm}e^{i\delta_u}|^2}\nonumber\\
&\simeq&
-2\frac{\Re [A_{u\kpm}e^{i(\delta_u-\delta_g)}]}{S_{g\kpm}},
\label{asym}
\ea
where $\delta_g$ and $\delta_u$ are scattering phases for gerade and ungerade states, respectively. 
Note that, according to \Eref{asym}, the asymmetry parameter $\beta>0$ corresponds to the case when the bare proton prefers to be ejected in the same direction with the photoelectron, while the H atom is ejected in the opposite
direction.

Since the matrix element $S_{g\kpm}$ is real, only the real part of
$A_{u\kpm}e^{i(\delta_u-\delta_g)}$ contributes to the asymmetry
parameter. The imaginary part of $A_{u\kpm}e^{i(\delta_u-\delta_g)}$
is the logarithmically divergent at the upper limit of the integral
over $t$ in \Eref{A_int_a}. This divergence appears in a situation
when the ejected electron interacts with the ion born directly in a
vibration continuum state. Asymptotically, the internuclear distance
in the ion in a vibration continuum state tends to $R(t)\to v_pt$
where $v_p=\kpm/m$ is the relative velocity of the nuclei. The
gerade and ungerade states are degenerate at large $R$. Following a
secular equation, in this case of the ion in an external field, the
nondegenerate right and left states are, respectively, 
$
\left|\text{r}\right\rangle=
(\left|1s\sigma_g\right\rangle+\left|2p\sigma_u\right\rangle)/\sqrt{2}
$
and 
$
\left|\text{l}\right\rangle=
(\left|1s\sigma_g\right\rangle-\left|2p\sigma_u\right\rangle)/\sqrt{2}
\ .
$
The left and right states describe the two halfs of the bound electron
cloud which is localized near the nuclei.  The difference of the
potential that the ejected electron affects on this half-clouds is
$2\mu=2d(R)/r_e^2=(\mathbf{n}_R\cdot\mathbf{v}_p/v_e^2)t^{-1}$. As the
result, the phase difference of the half-clouds is
$\sim(\mathbf{n}_R\cdot\mathbf{v}_p/v_e^2)\ln t$. The imaginary part
of $A_{u\kpm}e^{i(\delta_u-\delta_g)}$ is proportional to the limit of
this phase difference at $t\to\infty$. But, since the half-clouds do not
overlap for large $R$, this phase difference has no effect on any
observables.

\begin{figure}[ht] 
\includegraphics[angle=-90,width=0.9\columnwidth]{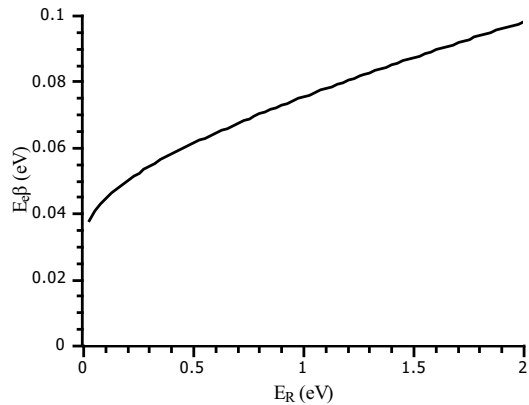}
\caption{The scaled asymmetry parameter $\tilde{\beta} = E_e\beta$ as
a function of $E_R$ for $\mathbf{R}\parallel\mathbf{n}_e$. }
\label{FIG_asymm}
\end{figure}
As follows from \Eref{A_int_a}, the asymmetry parameter is inversely
proportional to the ejected electron energy $E_e$. So we can introduce
the electron-energy-independent asymmetry parameter
$\tilde{\beta}(E_R)$ via
\be
 \beta = \frac{\tilde{\beta}(E_R)}{E_e}\ .
\label{beta_betatilde}
\ee 
In \Fref{FIG_asymm}, the parameter $\tilde{\beta} = E_e\beta$ is
shown as function of KER. To get an unscaled $\beta$ parameter for a
given ejected electron energy $E_e$, $\tilde{\beta}$ should be divided
on $E_e$ in eV.  
As is seen in \Fref{FIG_asymm}, for any KER the proton prefers to be
ejected in a direction that concides with the electron ejection
direction.  When the ejected electron energy is very small and
comparable with the excitation energy of the ion, the approximation
$r_e(t)=v_et$ is far too crude because the energy of the ejected
electron is changed noticeably after the excitation of the bound
electron. So, the condition of the validity of \Eref{beta_betatilde}
is $E_e\gg E_R$.


In conclusion, we propose a mechanism of the $p$--H symmetry
breaking in DPI of H$_2$ due to the \Hp ion interaction with the
ejected electron. 
A seemingly forbidden transition to the ionic $2p\sigma_u$ state at
low photon energies is enhanced by final-state interaction with the ionized electron due to the breakdown of the Frank-Condon principle.
This mechanism may be responsible for a noticeable asymmetry at low
KER.  It can be readily observed experimentally using a COLd Target
Recoil Ion Momentum Spectroscopy (COLTRIMS) coincident detection
technique and and a synchrotron light source. 
To resolve the dependence seen in \Fref{FIG_asymm}, an energy resolution of
the order of 0.1~eV is required. This gives an estimate of the
required photon bandwidth, the photoelectron energy and the KER
resolutions.  The modern COLTRIMS apparata are sensitive to asymmetry
down to 5\% \cite{Wu2013}.
For reaching of such an asymmetry, the energy of the ejected electron
should be the order of 1~eV.

Our estimates are based on the semi-classical approximation of the
ejected electron motion. A fully quantum-mechanical treatment is
needed to provide a more rigorous estimate for small values of the
ejected electron energies. This will require a solution of the fully
dimensional time-dependent Schr\"odinger equation with respect to the
electron and nuclear coordinates. This solution was sought and found
in \citet{Sansone2010} but the electronic part of the vibrational
states was calculated in a sphere of radius 160~a.u. This radius is
far too small to account for the effect under present
consideration. However, this approach should reveal the effect after
increasing this radius by several times.

The authors are thankful to Reinhard D\"orner for his interest in the
present work. One of the authors (VVS) wishes to thank the Australian
National University (ANU) for hospitality.  His stay at ANU was
supported by the Australian Research Council Discovery grant
DP120101805. VVS also acknowledges support from the Russian Foundation
for Basic Research (Grant No. 14-01-00420-a).


\end{document}